# On the Law of Directionality of Genome Evolution

Liaofu Luo

**Laboratory of Theoretical Biophysics, Faculty of Science and Technology,
Inner Mongolia University, Hohhot 010021, China**

*Email address: lolfcm@mail.imu.edu.cn

**Abstract**

The problem of the directionality of genome evolution is studied from the information-theoretic view. We propose that the function-coding information quantity of a genome always grows in the course of evolution through sequence duplication, expansion of code, and gene transfer between genomes. The function-coding information quantity of a genome consists of two parts, p-coding information quantity which encodes functional protein and n-coding information quantity which encodes other functional elements except amino acid sequence. The relation of the proposed law to the thermodynamic laws is indicated. The evolutionary trends of DNA sequences revealed by bioinformatics are investigated which afford further evidences on the evolutionary law. It is argued that the directionality of genome evolution comes from species competition adaptive to environment. An expression on the evolutionary rate of genome is proposed that the rate is a function of Darwin temperature (describing species competition) and fitness slope (describing adaptive landscape). Finally, the problem of directly experimental test on the evolutionary directionality is discussed briefly.

## 1   A law of genomic information

The traditional natural sciences mostly focused on matter and energy.  Of course, the two are basic categories in the nature. However, parallel to yet different from matter and energy, information constitutes the third fundamental category in natural sciences. One characteristic of a life system is that it contains a large amount of information. Life consists of matter and energy, but it is not just matter and energy. The life of an individual comes from the DNA of its parents. DNA, which weighs only $10^{-12}$ gram for human, is insignificant in terms of matter because, like many other things on earth, it is composed of nitrogen, oxygen, sulfur, etc. In addition, DNA, as a source of energy, is also unimportant, since it is just composed of the similar level of chemical energy as other macromolecules that can be produced by experiment. Schrodinger (1944) was the first who recognized the importance of information and indicated that the characteristic feature of life which differentiates from an inanimate piece of matter is the large amount of information contained in its chromosomes. He said, "We believe a gene – or perhaps the whole chromosome fibre – to be an aperiodic solid. With the molecular picture of gene it is no longer inconceivable that the miniature code should precisely correspond with a highly complicated and specified plan of development and should somehow contain the means to put it into operation." Seventy years have passed. Now, as we try to formulate the basic law of life system we should put our discussion based again on the concept of information. Any building or structure needs a blueprint to be built, this blueprint is the information. A life is a complicated structure. It is like a drama unfolding in a time span. The information that represents this dynamic structure is the soul of a life. DNA is a book of microscopic size that has all basic information about the development, growth and death of a life. This is also true when the point is examined in terms of the origination of life. Because life only comes into being when the primitive molecular sequence has evolved for a long time and has enough information accumulated in it. As we all admire the exquisiteness of the structure and function of advanced lives on earth, we know that they have all evolved from primitive lives. Each tiny bit of evolution must be realized through storing information in DNA and changing the DNA sequence. Therefore, whether we examine the issue from the emergence of individual life or from that of a species, life starts from information. Lao Tzu said in his *Book of Tao and Teh*, "It was from the Nameless that Heaven and Earth sprang, the named is but the mother that rears the ten thousand creatures, each after its kind." If the first sentence can be interpreted as: The Universe starts from nothingness, then the second one means that life starts from information.

Then, what is the basic law of genomic information? The extension of the diversity of species and their evolution towards higher function more adaptive to environment shows that the life evolution obeys a law with definite direction. Darwin expressed the law as "survival of the fittest". It means that the 'designed' properties of living things better adapted to survive will leave more offspring and automatically increase in frequency from one generation to the next, while the poorly adapted species will decrease in frequency. Here, the basic point of Darwin's theory is the directionality of the evolution. Accompanying the development of molecular biology and with ever increasing understanding on genomes we are able to express the law on life evolution direction more quantitatively and precisely. About the direction of genomic evolution many studies were carried out on the change of genome size (Gregory, 2005; Bennett et al, 2005; Gregory et al, 2007). The size of genomes for sibling species can change more than several tenfold, for example, 340-fold for flatworms, 70-fold for nematodes, 170-fold for arthropods (insecta), 350-fold for fish, 130-fold for amphibians,196-fold for algae (chlorophyta), 500-fold for pterridophytes and 1000-fold for angiosperms, etc. Moreover, the evolutionary complexity of a species is irrespective of its genome size. For example, the C-values of lungfishes are higher than human about 8 to 20 times. Some salamanders can also have large genomes with C-values 15 times or more than human. Next, about the genome size evolution it was speculated that plants might have a "one-way ticket to genomic obesity" through amplification of retrotransposons and polyploidy. The similar assumptions were also made that the animal genome sizes might change in the direction of increase. However, there has been considerable evidence that both increases and decreases may occur in plant and animal lineages. In the meantime, no fossil evidence on the genome size variability in the single direction was reported. These have made the issue on the directionality of genome size more complex and interesting. However, from our point of view the genome size of a species is not a proper measure of evolutionary directionality since it is irrespective of genetic information necessary for encoding the biological function. Instead, when our investigation is based on the function-coding information we will be able to obtain an unambiguous picture on the evolutionary law of genomes.

We propose that the Genome Evolution Direction obeys the law of Function-Coding Information Quantity Growing (CIQG).

Start from the definition of information quantity. For an n- long sequence (called sequence A) written by symbols $A_1$, $A_2$,….,and $A_n$ where $A_i$ taking $s_i$ possible values ($i$=1,…,n), if the sequence A encodes certain function then we define the function-coding information quantity of the sequence $I_C = \log_2 \prod_i s_i$.

There are three points about the definition of function-coding information quantity which should be clarified first. The first is related to the estimate of information quantity contributed by epigenetic inheritance, the second is how to estimate information gain in the process of inheritance information expression, and the third is to distinguish the function-coding part from the total information quantity of DNA sequence.

For DNA sequence of a genome the symbol $A_i$ takes four values A,G,C or T. However, for eukaryotes the chromatin remoulding and histone modifications is another type of variables which has the potential to influence fundamental biological processes and may be epigenetically inherited through somatic divisions (ENCODE, 2007). The nucleosome positioning, the occupancy of nucleosomes along the chromosome, as a key factor for gene regulation has been found in recent years. More than thirty histone modifications have been found for human and other vertebrates (Kouzarides, 2007). They form the source of heredity information in addition to DNA sequence of four kinds of bases A,G,C and T. Moreover, for prokaryotes and eukaryotes the DNA methylation can inhibit gene expression and be epigenetically inherited through perpetuation by a maintenance methylase and therefore the methylated bases give additional symbols applicable in genetic language (Lewin, 2008). Generally speaking, the epigenetic inheritance describes the ability of different states, which may have different phenotypic consequences, to be inherited without any change in the sequence of DNA. But in the long history of adaptive evolution the interaction between epigenetic inheritance factor and DNA sequence may lead to the connection between two factors, making the former dependent of the latter. The point is supported by some recent observations. For example, the nucleosome positioning both in human and yeast genome can be predicted from DNA sequence by using bioinformatics method (Ioshikhes et al, 2006; Gupta et al, 2008; Chen et al, 2010).

If the independent role of epigenetic inheritance in the information transmission can be neglected then the source of the genetic information can only be in DNA sequence. The next problem we are concerned is the information gain in DNA expression. DNA sequence is the main sequence of inheritance information. In its expression many protein sequences (called para-sequences) are formed under its instruction and these proteins are in close interaction with DNA sequence. For an $n$-long DNA sequence the primary information quantity is $2n$. Suppose the total length of protein sequences is $m$. Then the total information quantity of DNA-protein interacting system is $2n + m \log_2 20$. It means in the expression of genetic information the information quantity of a genome is largely increased. This is a universal principle that the information quantity increases in the expression and transmission due to formation of para-sequences and interaction with them. So, for a living cell the information quantity acquired in its life circle is much larger than the information quantity of primary inheritance information. However, in the proposed CIQG law the information gain from para-sequences is not taken into account and what we talk about is only the inheritance information of DNA main sequence.

What is the function-coding information quantity of DNA sequence? A function-coding DNA sequence consists of two basic types: p-coding sequence which encodes functional protein and n-coding sequence which encodes all other functional sequence segments. Evidently, the function-coding DNA sequence is a part of DNA main sequence. For a four-symbol DNA sequence the p-coding information quantity equals twice of the length of protein-coding sequence (or CDS in common genomics terminology), and the n-coding information quantity includes the contribution from promoters and other transcriptional regulatory elements upstream of the transcriptional start site (TSS) and the contribution from non-protein-coding RNAs (ncRNAs) in genes and intergenic sequences in primary transcripts. So, the function-coding information quantity of a genome

$$I_C = 2(N_p + N_n) \quad (N_p = \text{CDS length}, \ N_n = \text{n-coding sequence length}) \tag{1}$$

if the epigenetic inheritance effect is neglected. For most genomes $I_C$ is smaller than twice of genome size due to existence of non-functional part (for example, pseudogenes) in DNA sequence.

With the meaning of the function-coding information quantity $I_C$ we express the law of Function-Coding Information Quantity Growing (CIQG) as follows:

The function-coding information quantity $I_C$ (including protein-coding information quantity and other function-coding information quantity) of a genome sequence always grows in the course of evolution ($\frac{dI_C}{dt} \geq 0$) through sequence duplication, expansion of code, horizontal gene transfer and other expanding strategies.

Entropy increase is a universal law of nature. Due to the randomness of molecular motion the entropy of any isolated physical system always increases. However, the law stated above is not identical with the physical law of entropy growing. The law of CIQG refers to the evolution of function-coding information quantity of genome. The mechanism responsible for CIQG is functional selection. Those species that cannot increase their coding information quantity in genomic evolution should be deleted through species competition. As a biological law of evolution, there may exist exception; for example, some genome evolves in strange and peculiar environment. However, the exception should be rare. On the other hand, the time scale $dt$ in $\frac{dI_C}{dt} \geq 0$ is determined by the minimal time interval required for natural selection acting on heredity process. As using the clock with accuracy of several generations for given species the law $\frac{dI_C}{dt} \geq 0$ can then manifest itself.

The above proposal on the CIQG law is definitely different from the second law of thermodynamics of entropy growing, but it is consistent with Kauffman's hoped-for fourth law of thermodynamics for self-constructing systems of autonomous agents (Kauffman, 2000). Following Kauffman the autonomous agent means an agent or system working in one's own interest adaptive to environment. The genome discussed in this article can be looked as a concrete example of 'autonomous agent'. When our discussions are devoted for the system of genomes we are able to gain deeper insights on the basic evolutionary law on autonomous agents.

In his book "Investigation" Kauffman wrote: "Biospheres, as a secular trend, that is, over the long term, become as diverse as possible, literally expanding the diversity of what can happen next. In other words, biospheres expand their own dimensionality as rapidly, on average, as they can." and called it 'the fourth law of thermodynamics for self-constructing systems of autonomous agents'. He formalized the push into novelty as the mathematical concept of an adjacent possible.

"Biospheres may enter their adjacent possible as rapidly as they can sustain."

Through extensive and elaborate investigations we found many examples that show the genomes becoming as diverse as possible and expanding their own dimensionality continuously in the long term of evolution. Moreover, we suggested that the function-coding information quantity can be served as the quantity for measuring the diversity or dimensionality of the genome.

Supporting facts in the demonstration of CIQG law have been listed in another paper, Arxiv:0808.3323 (Luo, 2008), and the cited literatures therein. In next two sections we will discuss the driving forces of the directionality of genome evolution both from a genome sequence itself and from the competition among genomes. The evolutionary trends of DNA sequences revealed by bioinformatics will be studied in section 2. A theoretical formula on the genomic evolutionary rate will be proposed and the relation of evolutionary directionality with species competition will be discussed further in section 3.

## 2 Evolutionary trends of DNA sequences revealed by bioinformatics

**2.1 Gene sequence evolution obeys Maximum Information Principle.** By using Haken's Maximum Information Principle (MIP) formalism (Haken, 1988) we demonstrated that for the protein-coding sequence of gene the Shannon information quantity (information entropy)

$$H = -\sum_i p(i) \log_2 p(i) \tag{2}$$

($p(i)$ is probability of base $i$ in a given gene sequence) is maximized under the constraints of

$$N = \sum_i p(i) = 1$$

fixed Markov entropy $H_M$

$$H_M = -\sum p(i) F_i$$

$$F_i = \sum_j p(j|i) \log_2 p(j|i) \quad , \tag{3}$$

fixed Markov entropy with lag and/or fixed G+C content $p(C) + p(G)$. We found that the deduced maximized statistical distributions of $p(i)$ for most protein-coding sequences are in good agreement with experimental data (Luo, 1995). In the above deduction the maximization of information entropy describes the random mutation of bases in the gene sequence evolution; while the constraints of fixed Markov entropy or fixed G+C content describes the role of natural selection on particular base distribution and base pair correlation. Therefore, the driving force of sequence evolution is the combination of two factors, the random mutation and the natural selection. The success of MIP analysis shows the maximization of Shannon information quantity under constraints gives the evolutionary direction of gene sequences. This is micro-evolution. The evolution of genome is macro-evolution which results from a large amount of micro-evolution of genes (Ridley, 2004). The above discussion also gives an explanation of why the function-coding information quantity $I_C$ defined above is so important for describing the genome evolution since the information quantity (namely, the logarithm of the involved state number) is a basic characteristic of the information system .

**2.2 Functional sequence segment recognition by increment of diversity** Diversity of a sequence ensemble $X$ is defined by (Laxton, 1978)

$$D(X) = N \log_2 N - \sum_{i=1} n_i \log_2 n_i = N(-\sum_i p_i \log_2 p_i) \tag{4}$$

where $n_i$ the frequency of base $i$ in the ensemble and $p_i = \dfrac{n_i}{N} = \dfrac{n_i}{\sum_i n_i}$ its probability. Note that the diversity is exactly the Shannon information multiplied by $N$. For an ensemble of sequences of given function the base distribution of $\{n_i\}$ has a particular form which is determined by functional evolution. The diversity $D(X)$ provides a bioinformatics quantity for functional segment recognition. For sequence comparison between ensembles $X$ ($\{n_i\}$, $N$) and $Y$ ($\{m_i\}$, $M$) we can introduce the generalized Jensen-Shannon Divergence (Lin, 1991)

$$\text{GJS} = N \sum p_i \log \frac{p_i}{(Np_i + Mq_i)/(N+M)} + M \sum q_i \log \frac{q_i}{(Np_i + Mq_i)/(N+M)}$$

$$= N \sum p_i \log p_i + M \sum q_i \log q_i - (N+M) \sum s_i \log s_i \tag{5}$$

$$s_i = \frac{Np_i + Mq_i}{N+M}, \quad p_i = \frac{n_i}{N}, \quad q_i = \frac{m_i}{M}$$

GJS is also called Increment of Diversity

$$ID(X, Y) = D(X + Y) - D(X) - D(Y)$$

which means the increment of diversities as two sources $X$ and $Y$ merged. If $Y$ is a single sequence fragment whose function is to be recognized, and $ID(X, Y)$ is small enough then $Y$ is predicted as a sequence of same function with ensemble $X$. Using $ID$ method we have successfully recognized splicing sites, promoters and transcription starting sites, etc.(Lu et al, 2010) The success of functional segment recognition by Increment of Diversity means the bases in functional sequences have attained a stable distribution around the maximum entropy due to the functional selection in evolution (Jin et al, 2008). The success also shows the information quantity or diversity is a good quantity for functional sequence recognition.

**2.3 Information quantity gain in sequence recombination is correlated with recombination rate**    We study the increase of information quantity in sequence expansion. Consider the change of information quantity when two sequence segments are combined to one. The information quantity (diversity) of sequence segment $a$ (long $N_a$) or $b$ (long $N_b$) is

$$I(a) = -N_a \sum_j p_{aj} \log_2 p_{aj} = N_a \log_2 N_a - \sum_j m_{aj} \log_2 m_{aj}$$

$$I(b) = -N_b \sum_j p_{bj} \log_2 p_{bj} = N_b \log_2 N_b - \sum_j m_{bj} \log_2 m_{bj} \qquad (6)$$

respectively. The information quantity (diversity) of combined sequence is

$$I(a+b) = (N_a + N_b)\log_2(N_a + N_b) - \sum_j (m_{aj} + m_{bj})\log_2(m_{aj} + m_{bj}) \qquad (7)$$

It is easily to prove $I(a+b) \geq I(a) + I(b)$. Define the gain of information quantity in the combination

$$R(a;b) = \frac{I(a+b) - I(a) - I(b)}{I(a) + I(b)} \qquad (8)$$

For human genome taking the window width 5Mb ($N_a = N_b = 5M$) we found the averaged $R(a;b)$ is correlated to the experimental recombination rate (Liu et al, to be published). Since the recombination rate is a measure of local evolutionary rate of a species the above result means the genomic local evolutionary rate is related to the information gain in DNA sequence expansion.

**2.4    Deviation of k-mer frequency from random sequence increases with evolution**    For a long-$N$ DNA sequence the average of $k$-mer frequency is measured by α, $\alpha = \dfrac{N}{4^k}$, and the deviation by $\sigma_\alpha^2$, $\sigma_\alpha^2 = \sum_i \dfrac{(f_i - \alpha)^2}{4^k}$ ($f_i$ - the frequency of the $i$-th $k$-mer).    For a random sequence of same length, $\sigma_\alpha^2(k,N) = \alpha(k,N)$ due to Poisson distribution. Define the ratio of the deviation ($\sigma_\alpha^2(k,N)$) for a real DNA sequence to that for a random sequence ($\alpha(k,N)$) as the non-randomness parameter $Q(k,N)$

$$Q(k,N) = \frac{\sigma_\alpha^2(k,N)}{\alpha(k,N)} \qquad (\alpha = \frac{N}{4^k}) \qquad (9)$$

$Q(k,N)$ gives a sum rule of the k-mer frequency deviations which represents the nucleotide correlation strength or non-randomness level of DNA. It describes how the *k*-mer frequency distribution of a DNA sequence is deviated from the stochastic value due to functional selection. On the other hand, from the statistical analysis of genome data we found $Q(k,N)$ increases with *N* for most DNA sequences when *N* large. Since the sequence duplication is an important mechanism for genome evolution the above result shows the non-randomness increases due to sequence duplication apart from the functional selection. Therefore, both factors, the functionalization of DNA sequence (base mutation under selective constraint) and sequence expansion, are the driving force for genome evolution and both factors contribute to the increase of $Q(k,N)$.

The dependence of the sum rule $Q(k,N)$ on species has been calculated for *k*=6 and *N*=$N_f$ where $N_f$ is the length of genome, i.e. the total length of all chromosomes. The $Q(6,N_f)$ for 11 genomes are listed in Table 1 (Luo et al, 2010). We find it increases roughly with evolutionary complexity which means the directionality of genome evolution does exist.

In fact, by using *k*-mer frequency statistics one can reconstruct evolutionary tree of species. The evolutionary tree gives more detailed evolutionary relations among species. The deduced evolutionary tree from *k*-mer frequency statistics is basically consistent with life tree (Qi et al, 2004). However, in recent studies it was found apart from common evolutionary divergence the ecological convergence of the genomes is superimposed to the tree (Kirzhner et al, 2007). Both evolutionary divergence and ecological convergence are consistent with the above analysis based on the sum rule of k-mer frequency deviation.

Table 1    The relation between $Q(6,N_f)$ and species

| Species | $N_f$ | $Q(6,N_f)$ |
| --- | --- | --- |
| *M. genitalium* | 580076 | 244 |
| *A. fulgidus* | 2178400 | 212 |
| *E. coli* | 4639677 | 371 |
| *S. cerevisiae* | 12070900 | 1642 |
| *A. thaliana* | 118960610 | 23479 |
| *C. elegans* | 100269917 | 40371 |
| *D. melanogaster* | 72736145 | 7854 |
| *G. gallus* | 285755532 | 72776 |
| *D. rerio* | 859083401 | 133051 |
| *M. musculus* | 2617133082 | 389371 |
| *H. sapiens.* | 2893618752 | 494841 |

(Sequence data taken from http://www.ncbi.nlm.nih.gov/projects/genome/)

**2.5   Homogeneity of RNA folding energy in genome means coordinately functionalization of RNA local structure**    Luo and Jia ( 2007a, 2007b) calculated local mRNA folding energy in twenty-eight genomes. Through the local folding of mRNA sequence they discovered an interesting law: the intraspecific homogeneity and interspecific inhomogeneity of mRNA folding energy. Then, the studies were generalized to more genomes and to other functional regions of RNA sequences. The same results were obtained (data see below, next section). How to understand the intraspecific homogeneity of RNA folding in a genome? Although the large interspecific difference may occur due to the rapid accumulation of mutations the stochastic interaction among genes can make the horizontal spread of the change of RNA folding energy from one gene to other members. As is well known, the concerted evolution (coincidental evolution) of multigene families has been proved by a large body of data from restriction enzyme analysis and DNA sequencing techniques (Li, 1997). The intraspecific homogenization of mRNA folding energy is like the concerted evolution but may have alternative origin. We know that at each stage of the transfer of heredity information from primitive DNA sequence to functional protein expression the related elements of the genome are frequently interacted and re-organized. From sequence to structure to function, this is the basic logic of genome as a life machine. The homogenization of RNA folding energy among genes in a genome may reflect the close interaction between genomic sub-systems necessary for information transfer from sequence structure to function, reflect the functional need of the total genome.

The coordinately functionalization of RNA local structure is a key factor to fit the function of genome. To give a quantitative description of the functionalization of RNA local structure we introduce a fitness parameters of the structure: the non-stochasticity of RNA local folding. For gene $q$ in a genome the non-stochasticity parameter is defined by

$$D(q) = \frac{1}{s-1}\sum_{i}^{s}(E_i(q) - A(q))^2 \qquad (10)$$

where $E_i(q)$ is the $i$-th value of local folding energy of gene $q$ in a window of given width (say, 50 bp) and $A(q)$ is the local folding energy of the corresponding randomized sequence (preserving the same base composition as the native sequence) of gene $q$. Set the mean and the deviation of $D(q)$ denoted by $\mu$ and $\sigma$ respectively. Define $CV = \frac{\sigma}{\mu}$ to describe the scattering of $D(q)$ of a genome. Simultaneously, we calculate the k-moment of $D(q)$, $C_k = \langle (D(q)-\mu)^k \rangle$. Skewness is defined by $C_3/\sigma^3$ and Kurtosis defined by $\frac{C_4}{\sigma^4} - 3$. The small skewness and kurtosis means the distribution near Gaussian type. We study the relation between $D(q)$ and $q$. The linear regression gives the slop very near to zero, $\frac{dD(q)}{dq} \approx 0$, and the scattering of $D(q)$ is small for each genome. The parameters $CV$ and the slops of linear regression for 11 genomes are listed in Table 2. Skewness and Kurtosis are also listed (Chen, to be published). These results of statistical analyses show that the deviation of the local RNA folding energy from the corresponding stochastic value, $D(q)$, is indeed homogenized in a genome. Since RNA folding is an important step for the genetic information transmission and expression the homogeneity of RNA folding energy and of its deviation gives demonstration of the coordinately functionalization of RNA local structure to fit the functional need of the total genome.

Table 2  $D(q)$-related parameters for typical genomes

| Organism | Slope | CV | Skewness | Kurtosis |
|---|---|---|---|---|
| H. sapiens | -0.00385 | 0.376 | -0.143 | -0.412 |
| R. norvegicus | -0.00806 | 0.258 | -0.757 | 2.403 |
| D. rerio | -0.00191 | 0.346 | 0.583 | 1.853 |
| D. melanogaster | -0.00306 | 0.358 | 0.491 | 1.546 |
| C. elegans | -0.00194 | 0.271 | -0.168 | 0.983 |
| A. thaliana | 2.27E-5 | 0.391 | 0.548 | 0.587 |
| S. cerevisiae | 1.19 E-4 | 0.532 | 2.176 | 8.534 |
| B. subtilis | -5.22E-4 | 0.199 | 0.137 | -0.344 |
| E. coli | -1.45E-4 | 0.178 | 0.120 | 0.587 |
| A. fulgidus | -0.00133 | 0.209 | 0.344 | 1.321 |
| M. pneumoniae | -0.00263 | 0.168 | 0.159 | 0.372 |

**2.6 Evidence of function differentiation among species : the large interspecific difference of RNA folding energy**   The first round bioinformatics studies on the RNA function differentiation were carried out by Luo and Jia in 2005 to 2006 (Luo et al, 2007a, 2007b). They discovered the intraspecific homogeneity and interspecific inhomogeneity of mRNA folding by calculating local mRNA folding energy in twenty-eight genomes, 8 archaea, 14 Eubacteria and 6 eukaryota.  For each species about 120 genes (coding regions) were chosen stochastically and each mRNA sequence was folded in a local window pattern, namely, in short regions of 50 bases and shifted by 10 bases.   The averaged local folding energy of the *j*-th native mRNA sequence in the *i*-th genome is denoted by $y_{ij}$. Define the mean error sum of squares MESS and the mean class sum of squares MCSS, the former describing the square deviation of mRNA folding energy in a genome and the latter describing the square deviation of this energy among genomes.   The analysis of variance (ANOVA) is given by the following formulas:

$$MESS = \frac{SSe}{\sum_{i} n_i - t}$$

$$SSe = \sum_{ij}(y_{ij} - \overline{y_{i\cdot}})^2 \qquad (11)$$

and

$$MCSS = \frac{CLS}{t-1}$$

$$CLS = \sum_{ij}(y_{ij}-\overline{y})^2 - \sum_{ij}(y_{ij}-\overline{y_{i.}})^2 \qquad (12)$$

Here *SSe* is the error sum of squares, *CLS* is the class sum of squares, *t* is the number of organisms in a given class, $\overline{y_{i.}}$ is the mean of $y_{ij}$ over the $n_i$ sequences in the *i*-th organism for a given class ($i=1,2,...,t$; $j=1,2,..,n_i$). If $y_{ij}$ depend on another set of co-variables $x_{ij}$ and these two sets of variables obey a regression equation in the form of $y_{ij} \approx a+bx_{ij}$ for each class then one should generalize ANOVA to the analysis of covariance (ANCOVA). Through ANOVA and ANCOVA analysis these authors calculated MESS and MCSS of folding free energy of native mRNA and deduced the *F* value

$$F = \frac{MCSS}{MESS} \qquad (13)$$

By comparing the calculated *F* value with *F*-distribution the significance level (P-value) was deduced. They discovered $F \gg 1$, i.e., the extremely significant difference between MCSS and MESS.

In the above studies only coding RNA sequences and mainly prokaryotic genomes were considered. To explore if the intraspecific homogeneity and interspecific inhomogeneity of mRNA folding is universal Chen et al generalized the statistical analyses to more eukaryotic genomes and for more functional regions in addition to coding sequences (Chen, to be published). Consider 32 organisms including 7 mammalian (class 1), 7 other vertebrates (class 2), 6 invertebrates (class 3), 8 plants (class 4) and 4 fungi (class5). For each genome 100 genes were stochastically chosen and four types of genomic regions, namely 5′ UTR, intron, CDSs and 3′ UTR, are studied. The statistical property of mRNA folding free energies was studied by use of two methods, the one-way analysis of variance (ANOVA) and the analysis of covariance (ANCOVA). In ANOVA the co-variables $x_{ij}$ are not introduced and the mean error sum of squares *MESS* and the mean class sum of squares *MCSS* are calculated directly from $y_{ij}$. While in ANCOVA only the simple model is adopted where the co-variable is supposed to be $x_{ij}=(G+C)\%$ for 4 different functional regions respectively. The results of ANOVA analysis on RNA folding energy for 5 classes of organisms are listed in Table 3, which gives $F = 10\sim50$ at the significance level $P<0.0001$. Furthermore, if 32 organisms are put into a single class then the like calculation gives

$F_{5'UTR}=37.16$, $F_{CDSs}=32.79$, $F_{Intron}=35.56$, $F_{3'UTR}=15.61$ with $p<0.0001$.

The similar results were also obtained in ANCOVA analysis on RNA folding energy (data not shown here). All above analyses indicate that the selection for RNA folding is specific among all genomes, irrespective of the kingdoms of species and the functional regions of genome. These results show the extremely significant difference between MCSS and MESS for folding energy of native RNA sequence is a universal law. Further, by use of the same method, Luo et al (2007a, 2007b)) and Chen et al (to be published) proved that the extremely significant difference between MCSS and MESS also exists for folding energy deviation of native RNA relative to randomized sequence.

What is the mechanism responsible for the universal law that the RNA folding energy (or its deviation from randomized sequence) in different genomes is always higher than that in a genome ? We speculate that it is a result of functional differentiation among species since more similar forms will compete more strong, which tends to push species apart during evolution （"divergence of churacter" as stated by Darwin）. So, the above calculations show that the Darwin's principle of evolutionary divergence can be demonstrated quantitatively by genome sequence analysis. In the mean time, the ecological convergence of the evolutionary tree proposed by recent studies (Kirzhner et al, 2007) can also be obtained in above RNA folding energy analysis among genomes. In fact, although the large $F$ values have been deduced in ANOVA and ANCOVA statistical analysis for prokaryotes and large branches of eukaryotes, the multiple comparison test shows the difference of folding energy deviation is not so significant for several particular pairs of genomes. So, both evolutionary divergence and ecological convergence are consistent with and can be deeply understood through RNA folding energy analysis. Our results show that the specific selection among genomes does exist from the point of RNA structure and function, and the functional selection among genomes is due to species competition adaptive to environment (including ecological environment) that serves as one of the firmest grounds for the understanding of the genomic evolutionary direction.

**Table 3   ANOVA analysis of RNA folding energy in 32 eukaryotes**

|  | $F(df_1, df_2)$[$] | | | |
| --- | --- | --- | --- | --- |
|  | 5′-UTR | INTRON | CDS | 3′-UTR |
| Class1 | 16.08 (6,693)* | 56.32 (6,693)* | 17.83 (6,693)* | 8.24 (6,693)* |
| Class2 | 22.15 (6,693)* | 45.78 (6,693)* | 8.63 (6,693)* | 26.11 (6,693)* |
| Class3 | 29.61 (5,594)* | 65.24 (5,594)* | 46.31 (5,594)* | 20.88 (5,594)* |
| Class4 | 39.66 (7,792)* | 37.48 (7,792)* | 40.98 (7,792)* | 19.21 (7,792)* |
| Class5 | 13.38 (3,396)* | - | 48.72 (3,396)* | 27.13 (3,396)* |

[$] $df$ represents the degree of freedom. $df_1 = t-1$, $df_2 = \sum n_i - t$. * means $p<0.0001$. Since most of genes in the fungi genome are intronless, comparisons related to intron regions were only carried out for class 1 to 4.

**To conclude:** 1) Firstly, both Shannon information quantity and diversity are good quantity for describing gene evolution. Both the maximization of Shannon information with respect to stochastic mutation and the increase of diversity with respect to sequence expansion imply the directionality of evolution. 2) Secondly, the functional selection results in the deviation of DNA segments from random sequences. The stronger the functional selection is, the larger the deviation from random will be. So, the functional constraints should be imposed to the maximization of information quantity. The random mutation, the sequence duplication and the functional selection are three independent driving forces but they jointly play a role in the genome evolution. This is the very reason why in our definition of the information quantity of genome (namely, CIQ) the 'function-coding' has been emphatically indicated. 3) Thirdly, the homogeneity of RNA folding energy in a genome and the significant difference of RNA folding energy among genomes imply the existence of the coordinately functionalization of RNA local structure (functional specificity) and the functional selection upon species competition adaptive to environment, that provides an important approach to the understanding of evolutionary direction of genome. To summarize, the genome evolutionary direction is generally decided by two factors, the inherent evolutionary trends of DNA sequence in a genome on one side, and the forces from outer, namely from species competition adaptive to environment on the other side. Here the inherent evolutionary trends include the evolutionary stochasticity of DNA sequence, the recognition of different functional segments by particular base distribution, the sequence recombination rate correlated with the gain of information quantity，the increase of non-randomness of k-mer frequency distribution to adapt to the functional need, and the homogeneity of RNA folding energy in a genome, etc. The microevolution is the basis of macroevolution. The consistency between microevolution and macroevolution is one of the firmest buttresses of modern evolutionary theory. What we learn about microevolution in this section is relevant to studying macroevolution, to solving the problem of genomic evolutionary direction.

## 3  Further discussions on genome evolutionary rate and species competition adaptive to environment

**3.1  Evolutionary rate of genome**   The species competition and functional selection is one of the most important mechanisms for function-coding information quantity growing of genome. The experimental evidences on CIQG law were discussed in Arxiv:0808.3323 (Luo, 2008). Several points related to the evolutionary rate, species competition, environmental change and experimental test of CIQG law will be discussed in the last sections. To describe the competition strength among genomes in a given ecological and inorganic natural environment one introduces the concept of Darwin temperature $T_D$. $T_D=0$ means no competition at all. A small $T_D$ means the relaxing competition and a high $T_D$ means the intense competition. The genome evolutionary rate $\frac{dI_C}{dt}$ is dependent of Darwin temperature. In mathematical terminology $\frac{dI_C}{dt}$ is a growing function of $T_D$. On the other hand, the evolutionary rate depends on the adaptation of the species. Suppose a set of fitness functions $\{w_\alpha(q)\}$ defined on sequence space $q$. Here $\{w_\alpha\}$ includes the characteristics by which taxonomists distinguish related species, it also includes some invisible adaptations such as disease resistance.   If the fitness changes suddenly in adaptive landscape, or in other words, if the slope of fitness $\{\frac{\partial w_\alpha}{\partial q_\beta}\}$ is steep enough, then the evolutionary rate will be high; otherwise, the small slope corresponds to low evolutionary rate. We assume $\frac{dI_C}{dt}$ is a function of the slope of fitness $\{\frac{\partial w_\alpha}{\partial q_\beta}\}$ and $T_D$,

$$\frac{dI_C}{dt} = F(\{\frac{\partial w_\alpha}{\partial q_\beta}\}, T_D) \qquad (14)$$

Note that due to the multi-dimensionality of fitness $\{w_\alpha(q)\}$ and the existence of correlation between them the adaptation of a species is not a single-factor quantity that can be defined properly. Haldane wondered why evolution so often seems nonadaptive or even maladaptive. For example, he remarked on the apparent nonadaptiveness of several basic characteristics of some species; he noticed that many lineages evolved large size or large horn just prior to extinction, etc. He went to some trouble to explain how this could be so (Haldane, 1990). However, from our point of view, the key problem is the multi-dimensionality of fitness function which should be noticed carefully in finding a reasonable explanation for this puzzle. With the same reason the slope of fitness $\frac{\partial w_\alpha}{\partial q_\beta}$ is a quantity of high dimension and $\frac{dI_C}{dt}$ may have complicate relation with the fitness slopes on the adaptive landscape.

To simplify the problem consider only one component is dominant in fitness functions and other components can be neglected. In this case one may assume the most probable trajectory on adaptive landscape is that with the steepest slope due to natural selection. Set

$$\underset{\beta}{Max}\{\frac{\partial w}{\partial q_\beta}\} = S_A \qquad (15)$$

By appropriate definition of Darwin temperature we speculate a simple form of $\frac{dI_C}{dt}$ (time $t$ in generation）

$$\frac{dI_C}{dt} = kT_D S_A \qquad (16)$$

It means for species in a given ecosystem, the evolutionary rate of genome is proportional to the ruggedness (steepness) of adaptive landscape and the proportional coefficient is a measure for the Darwin temperature (the competition intensity) of the ecosystem. $S_A$ in Eq (16) is species-dependent. For example, as a new species bifurcated from the evolutionary tree the new and old species have different adaptation slopes, $S_A$(old) $\neq S_A$(new). Generally, the new species has steeper adaptation slope and evolves faster than the old one. This is consistent with the fossil records that the evolution always shows a sudden rate in new species formation. Accompanying the function innovation and improvement the new species can have a higher function-coding information quantity and attain a higher evolutionary complexity.

For a species at the local maximum of fitness on adaptive landscape where all trajectories have zero slope, it leads to $\frac{dI_C}{dt}=0$. From Eq (16) the system simply moves to the nearest adaptive peak and remain there. However, the practical situation may be somewhat complicated. In above formulation the fluctuation force has been completely neglected. If the fluctuation force is added at RHS of Eq (14) and (16) such strict deterministic behavior does not occur and the random drift can move the genome under the control of a higher adaptive peak. In this way a succession of peaks can be reached, each one higher than the previous one. Another factor to drive the system not captured at local maximum is the environmental change since the adaptive landscape changes with environment. In any case, the evolution at pause is consistent with the punctuated equilibrium observed in the fossil records of species evolution (Ridley, 2004).

**3.2 Species competition and parasitism** The intensity of species competition in a ecosystem is described by Darwin temperature. The evolutionary directionality exists only in case of $T_D>0$. In the limiting case of no competition, $T_D=0$, the evolutionary rate approaches to zero and the directionality has no meaning. The CIQG law holds for all free-living genomes. Is it valid for parasites and symbionts? For parasite system the Darwin temperature needs re-definition. Consider a prokaryote genome varying from free-living to host-dependent lifestyle. The environment inside the host cell contains many of the nutrients and defense systems that a bacterial cell needed, since genes that are needed in a free-living bacterium to provide the resources are not needed in an intracellular bacterium. In this case the gene loss may be advantageous since a cell with less DNA can reproduce faster. The loss of function in parasitism results in the decrease of coding information quantity of the genome. This is a phenomenon of so-called retrogression, which should not be included in the scope of CIQG law. As for the evolution of bacteria parasitized in a stable host environment whether the CIQG law holds or not is a new problem to be studied. For example, in case of parasitic bacteria resisting drugs the evolutionary rate of the bacterial genome may be high and may not follow the law assumed above. Different from free-living bacteria the parasite uses some genetic apparatus (for example, the translational apparatus) of host for its own purpose. So, the parasite's genome is not an independent autonomous agent and its evolutionary law may adopt a form different from CIQG law.

**3.3 Sudden change of environment and evolution of survival** The survival of the fittest means the golden judge for the species competition is the adaptation to the environment. As environments change, the competition between similar individuals will make for the evolution of new adaptations in each that reduce the intensity of competition. So, the law of evolutionary direction is closely related to environmental change. For a genome in stable environment $I_C$ grows with time. For a genome in slow-varying environment $I_C$ grows adiabatically adapting to the change of environment. Here slow-varying means the marked environmental change can be found only in several generations of the species. However, for a genome in suddenly-changing environment, if the speed of growing $I_C$ cannot adapt to the sudden change of environment then the species would be close to extinction. Generally speaking, the sudden change of environment causes deaths of most individuals. Perhaps, only few quasi-species survive in the disaster. The survival may be explained by the quasi-species having evolved a structure of preadaptation which can evolve new function to adapt to the suddenly changing environment. The preadaptation is a common phenomenon observed in the evolution of species. The occurrence of the structure of preadaptation is an evidence of the genome as an autonomous agent always entering its adjacent possible as rapidly as it can sustain.

Is the CIQG law consistent with DNA damage? Consider bacteria live in an environment where the strong electromagnetic field is suddenly applied. Then the bacterial DNA could be damaged. DNA damage means the loss of heredity information. Although the decrease of genome information quantity does occur in this case it is merely due to external forces and is irrelevant to any evolutionary law. What is important is in the subsequent repair process, although a part of functional elements in old species may have been deleted but they have been replaced by a more efficient functional network. The process of evolving new function does obey the law of CIQG. Thus the increase of coding information quantity can be valid even in the circumstance of DNA damage.

**3.4 On experimental test of CIQG law** To test the CIQG law by direct experiments we suggest to study the evolution of a simple system – the free-living bacteria under external forces, for example, under ion implantation. Recently the bacterial genomic evolution was studied by Barrick et al (2009) through 40,000 generations from a laboratory population of *E coli*. The long-term experiment was conducted in 20 years. They observed the mutations accumulated at a near-constant rate over the first 20,000 generations and the rate of genomic evolution accelerated markedly only when a mutator lineage became established by about generation 26,500. It seems that the time needed for the occurrence of a mutator in free-living bacteria is still too long. So, we consider external forces applied to the prokaryotic genome system. Since the biological effects of low energy ion beam implantation on rice were discovered by Chinese scientist in 1986, the ion implantation as a new mutation source has been widely applied to the breeding of crops and microorganisms (Song et al, 2006). The basic researches on ion-beam-induced biological effects were also conducted in recent years. From the experiments on low energy (about 10 kev) ion beam implantation on *E coli* we observed the accelerated evolutionary rate of genome. Although most bacteria died under the irradiation of 10 kev ion beam the mutator phenotypes of *E coli* seem to appear in the survivals several days after ion implantation. In the mean time, if put the sample under periodic irradiations of ion beam (ion implantation repeated several times in a month), we found the genomic evolution still in one-way directionality, not showing any periodicity in spite of the environment having changed periodically. The preliminary result reveals that the experiments by using external field (electro-magnetic field, ion beam irradiation, etc) applied to organisms could provide an efficient tool to observe the genetic change of genome and test the CIQG law directly.

**Acknowledgement**
The author is indebted to Drs Wei Chen, Yang Gao, Jun Lu, Guoqing Liu and Zhiqin Song for their helpful discussions. The main numerical results given in Table 2 and 3 were calculated under Chen's help and the numerical results in Table 1 were originally obtained by Gao in a published article, arXiv: 1004.3843.

## References

Barrick JE, Yu DS, Yoon SH, Jeong H, Oh TK, Schneider D, Lenski RE, Kim JF. Genome evolution and adaptation in a long-term experiment with *Escherichia coli*. Nature 04080 doi:10.1038 (2009)

Bennett MD, Leitch IJ. Genome size evolution in plants. In: *Evolution of the Genome* (Edi. Gregory TR), Elsevier Inc 2005.

Chen W, Luo LF, Zhang LR. The organization of nucleosomes around splice sites. Nucl Acid Res 38: 2788-2798 (2010)

Chen W, Luo LF.  RNA Folding Energy: Homogeneity driven by coordinately functioning and specific selection among genomes ( to be published).

ENCODE Project Consortium. Identification and analysis of functional elements in1% of the human genome by the ENCODE pilot project. Nature 447:799-816 (2007).

Gregory TR. Genome size evolution in animals. In: *Evolution of the Genome* (Edi. Gregory TR), Elsevier Inc 2005.

Gregory TR et al. :Eukaryotic genome size databases. Nucleic Acids Research **35,** Database issue D332-D338 (2007).

Gupta S, Dennis J, Thurman RE, Kingston R, Stamatoyannopoulos JA, Noble WS.  Predicting human nucleosome occupancy from primary sequence. PLoS Comput Biol. 4:1-11 (2008).

Haldane JBS. *The Cause of Evolution* with introduction by EG Leigh Jr. Princeton University Press, Princeton, New Jersey, 1990.

Haken, H. *Information and Self-organization*. Berlin:Springer-Verlag 1988.

Ioshikhes IP, Albert I, Zanton SJ, Pugh BF. Nucleosome positions predicted through comparative genomics. Nat Genet, 38:1210~1215 (2006).

Jin HY, Luo LF, Zhang LR.  Using estimative reaction free energy to predict splice sites and their flanking competitors. Gene 424:115-124 (2008).

Kauffman S. *Investigations*. Oxford University Press, Inc, 2000.

Kirzhner V, Paz A, Volkovich Z, Nevo E, Korol A.  Different clustering of genomes across life using the T-T-C-G and degenerate R-Y alphabets: Early and late signaling on genome evolution? J Mol Evol. 64:448-456 (2007)

Kouzarides T. Chromatin modifications and their function. Cell, 128**:** 693-705 (2007).

Laxton RR. The measure of diversity. *J Theor Biol*. 1978;70(1): 51-67.

Lewin, B. *Gene VIII*.  Pearson Education Inc., 2004;  *Gene IX*. Jones & Bartlet Publishers, Inc.,2008.

Lin J   Divergence measures based on the Shannon entropy. *IEEE T Inform Theory* 37:145–151 (1991).

Li WH. *Molecular Evolution.* Massachusetts: Sinauer Associates 1997.

Liu GQ, Luo LF. The growing rate of the information quantity of the human genome is mediated by recombination (to be published).

Lu J, Luo LF, Zhang LR, Chen W, Zhang Y. Increment of diversity with quadratic discriminant analysis – an efficient tool for sequence pattern recognition in bioinformatics. Open Access Bioinformatics 2:89-96 (2010)

Luo LF, Jia MW. Messenger RNA information: Its implication in protein structure determination and others. In：*Networks: From Biology to Theory* (Edi: Feng, Jost & Qian) pp 291-308. London: Springer 2007.

Luo LF, Jia MW. Messenger RNA relating to protein structure.. In：*Leading Edge Messenger RNA Research Communications* (Edi: MH Ostrovskiy) pp 67-78. New York: Nova Science Publishers 2007.

Luo LF. Law of genome evolution direction: Coding information quantity grows. arXiv: 0808.3323 (Quantitative Biology 2008), http://arxiv.org/abs/0808.3323 ;
Front. Phys. China 2009, 4:241-251. DOI 10.1007/s11467-009-0014-x.

Luo LF, Gao Y, Lu J. Information-theoretic view of sequence organization in a genome. arXiv: 1004.3843 (Quantitative Biology 2010), http://arxiv.org/abs/1004.3843

Qi J, Wang B, Hao BL . Whole proteome prokaryote phylogeny without sequence alignment: a K-string composition approach. J. Mol. Evol., 58, 1–11 (2004).

Ridley M. *Evolution* (3rd edition). Blackwell Publishing 2004.

Schrodinger E. *What is Life*?  Cambridge : Cambridge Univ. Press.1944.

Song ZQ, Liang YZ, Zhang XS, Luo LF. Biological effects of low energy ion beam implantation on plant. Current Topics in Plant Biology 7:75-84 (2006).